\documentclass[useAMS,usenatbib]{mn2e}
\usepackage{graphicx}
\usepackage{epsfig}
\def\lesssim{\la}
\def\gtrsim{\ga}

\title[]{Capture of Irregular Satellites via Binary Planetesimal Exchange Reactions in Migrating Planetary Systems}
\author[]{Alice C. Quillen, Imran Hasan, \& Alex Moore  \\
{Department of Physics and Astronomy, University of Rochester, Rochester, NY 14627, USA; 
}  \\
}

\begin{document}
\label{firstpage}
\maketitle

\begin{abstract}

By logging encounters between planetesimals and planets 
 we compute the distribution of encounters  
in a numerically integrated two planet system that is migrating due to interactions with an exterior planetesimal belt.
Capture of an irregular satellite in orbit about a planet through an exchange reaction 
with a binary planetesimal is only likely when the binary planetesimal undergoes a slow and close encounter
with the planet.     
In our simulations we find that close and slow encounters between planetesimals and a planet
primarily occur with the outermost and not innermost planet.
Taking care to consider where a planet orbit crossing binary planetesimal would first be tidally disrupted, 
we estimate the probability of both tidal disruption and irregular satellite capture.
We estimate that the probability that the secondary of a binary planetesimal is captured and becomes
an irregular satellite about a Neptune mass outer planet is about 1/100 for binaries with masses and separations similar to transneptunian planetesimal binaries.   
If young exoplanetary debris disks host a binary planetesimal population then 
outwards migrating outer planets should host captured irregular satellite populations.
We discuss interpretation of emission associated with the exoplanet Fomalhaut b 
 in terms of collisional evolution of a captured irregular satellite population that is replenished due
 to planetary migration.
 
 
\end{abstract}

\section{Introduction}

The majority of irregular satellites about the gas giant planets in our Solar system
are in eccentric and high inclination orbits  suggesting that likely
originated in heliocentric orbit and were later captured into orbit within a planet's Hill sphere
(see the review by \citealt{jewitt07}).  Irregular satellite populations could have at one time been a source
of heavy dust production \citep{bottke10}.  If irregular satellite populations exist in distant planetary systems
 then exoplanets could be discovered by  detecting 
 scattered stellar light from their associated dust clouds \citep{kennedy11}.   
The color of the visible light from the detected object, Fomalhaut b,  in orbit about the nearby star 
Fomalhaut, is consistent with that of reflected
light from the star \citep{kalas08} and so does not originate from the planet itself but could arise
from a planetary ring system (as discussed by \citealt{kalas08}),  dust cloud associated with
irregular satellites (as proposed
by \citealt{kennedy11}) or circumplanetary disk akin to the birth site of
the Galilean satellites (e.g., as explored by \citealt{canup02}).

We focus on extrasolar planets that are in proximity to  a debris disk as this disk could host 
 planetesimals that could be captured by the planets in the system,  becoming irregular
satellites. 
Neither the Fomalhaut system or the HR8799 system are likely to have strongly scattered or thick
planetesimal disks, the Fomalhaut system \citep{kalas05,kalas08} because the disk is thin (with disk
aspect ratio only $h/r \sim 0.013$ \citep{quillen07}, and the HR8799 system \citep{marois10} because
the planets are closely spaced and the time until the planet's cross each
others orbits  is probably short \citealt{god09,fabrycky10,moore11}) implying that the planets have not
yet experienced close encounters.
The Fomalhaut and HR8799 systems can be considered to be analogs to our solar system prior
to the late heavy bombardment epoch of our solar system that involved close approaches
between the giant planets (the `Nice' model; \citealt{nice}) and subsequent scattering of the planetesimal disk.

In our Solar system,
the irregular satellites were likely captured during planet/planet close approaches during the
late heavy bombardment era  \citep{nes07}. 
However, the recent discovery of a large transneptunian binary population \citep{noll08} has motivated
study of the exchange reaction mechanism for irregular satellite capture 
\citep{agnor06,vok08,philpott10,nogueira11}.
Irregular satellites in orbit around the giant planets prior to the epoch of late-heavy bombardment
were likely lost during close approaches between planets \citep{nes07}.   
Little is known about possible irregular satellite populations that could have existed prior to the late
heavy bombardment era.

Both exoplanets Fomalhaut b and HR8799 b are located just interior to dusty disks that are generated from
collisions of planetesimals and so are called debris disks \citep{marsh05,su09}.
Planetesimals originating from these disks can be scattered by the outer planet 
into orbits that cross the orbits of interior planets.  
The interaction between planetesimals and planets allows the outer planet to migrate outwards into
the planetesimal disk via the exchange of angular momentum between planets
and planetesimals \citep{fip84,ida00}.  As the outermost planet migrates outwards the population  
of planet crossing planetesimals can be continually replenished.

 \citet{kennedy11} showed
that a dust producing irregular satellite population could be long lived, however they did
not explore mechanisms for irregular satellite capture about exoplanets.  
If the planetesimal disks in these systems contain  binary planetesimals, then the exchange
reaction mechanism for irregular satellite capture \citep{agnor06} is a viable mechanism for the capture of irregular
satellites about planets from the population of planet-orbit crossing planetesimals.
Currently, it is estimated that binaries account for 30\% of transneptunian or Kuiper belt objects (KBOs) 
with inclinations $<5^\circ$,  and $\sim5$\% of the rest of the KBOs \citep{noll08}.   
If binary planetesimals are common in extrasolar planetesimal  
disks then the migration of an outer planet into a planetesimal disk 
would cause binary planetesimals to cross the orbits
of giant planets in the system, facilitating the capture of irregular satellites.

A planetesimal binary is disrupted inside the Hill sphere of a planet when the distance to the planet
is sufficiently small that the planet's tidal force overcomes the gravitational attraction of the binary.  
The disruption can leave an object in a bound orbit around the planet  if the velocity of the binary components is sufficiently large 
and oriented so as to decrease the energy of the lower mass planetesimal when the binary is disrupted 
\citep{agnor06,vok08}. 
The probability 
of irregular satellite capture via binary exchanges depends both on the statistics of close approaches and their energies \citep{vok08,philpott10,nogueira11}.
In this paper we consider the statistics of close approaches between planetesimals and planets
in systems containing two migrating planets.   Our goal is to better understand mechanisms for 
irregular satellite capture that might operate
in young exoplanetary systems and prior to the epoch of late heavy bombardment in our Solar system.



\section{Close encounter distribution for a migrating two planet system}

We numerically study systems containing two planets that are  in proximity to a planetesimal disk that is exterior
to the two planets.  We restrict our study to low eccentricity and low inclination systems.
We first discuss our numerical integrations.  We compute the distribution of the first close encounter
that would disrupt a planetesimal binary.   We then consider the velocity distribution of these encounters.

\subsection{Numerical Integrations}

We have modified the hybrid symplectic integrator QYMSYM \citep{moore11} to record properties of all particles
that approach within a Hill radius of a planet.   The integrator runs at two levels, an outer level
where all particles are integrated with a symplectic integrator with a fixed timestep, $\tau$.  
When a particle, or particles pass within a planet's Hill radius
 the encounter is integrated more carefully with an adaptive stepsize N-body integrator.
Here  we use the word encounter to denote a trajectory that passes within the Hill radius of a planet,
 not a collision with a planet.
For every particle that passes within a planet's Hill radius  
in each larger timestep, $\tau$, we use the adaptive step-size finer integration to 
record the minimum distance to the planet, and at that location we record the relative energy per unit mass 
\begin{equation}
E_p = {({\bf v - v}_p)^2 \over 2} - {GM_p \over |{\bf r - r}_p|},
\end{equation}
where ${\bf v},  {\bf v}_p$ are the velocities of particle and planet, ${\bf r}, {\bf r}_p$ are the positions of
particle and planet, respectively, and  $M_p$ is the mass of the planet.
For $E_p > 0$ we compute a velocity 
\begin{equation} 
V_\infty \equiv \sqrt{2E_p}
\label{eqn:vinf}
\end{equation}
corresponding to the velocity
of the particle distant from the planet and with respect to the planet, were the particle and 
planet isolated and not orbiting the central
star.  This velocity can be used to estimate the probability that an exchange
reaction with a binary planetesimal can occur leaving behind a bound satellite (e.g, \citealt{vok08}) though
a more accurate calculation uses the Jacobi constant \citep{philpott10} or directly integrates orbits of all four bodies \citep{nogueira11}.   Here we do not integrate the four body problem for 
 a planetesimal  binary moving in the gravitational field
of a planet in orbit about a star. However we do integrate the trajectories of single planetesimals in the gravitational field a planet in orbit about a star and using these trajectories we estimate
the probability that a binary would disrupt near the planet. 

We work in units of the mass of the central star $M_*$, with distance given in units of the initial semi-major
axis of the innermost planet, $a_1$, and time in units of the initial orbital period of the innermost planet.
The masses of the planets are $M_1,M_2$ and semi-major axes $a_1,a_2$.  
The initial semi-major axes of the planets are $a_1 = 1$ and $a_2 = 1.4$. The initial planet inclinations
and eccentricities were set to zero.   The initial mean anomaly of the two planets were randomly chosen.

The planetesimal disk properties are identical for all integrations.
The initial planetesimal disk is comprised of $N=8192$ objects of mass $m=10^{-7}$ that are distributed
between semi-major axes of $a_{min} = 1.6$ and  $a_{max}=2.5$. The distribution of  planetesimal semi-major axes
is flat with probability independent of $a$ within $a_{min}$ and $a_{max}$.  The initial eccentricity and inclination distributions were chosen using  Rayleigh distributions
with the mean eccentricity $\bar e$ equivalent to twice the mean value of the inclination $\bar i$
and $\bar i = 0.01$.  The initial orbital angles (mean anomalies, longitudes of pericenter and longitudes
of the ascending node) were randomly chosen. 
The total mass of the disk
is nearly $10^{-3}$ and large enough that the outer planet can migrate a substantial distance through
the disk in a few thousand orbits making it feasible to run a number of similar integrations.

Our integrations primarily have inner planet with $M_1 = 10^{-3}$ or $M_1 = 10^{-4}$, however a few
were done with intermediate or larger masses.   
The properties of the numerical integrations are summarized in Table \ref{tab:tab1}.  
 Each simulation was integrated for at least $P=1900$ orbits with orbital periods are measured in units of 
 the inner planet's initial orbital period.
The outer planet migrated outwards in all simulations, however, in some simulations 
(those with low mass outer planets) the outer planet
first migrated outwards and then reversed direction.
   As a consequence we consider the statistics of encounters
only during the portion of the integration when the outer planet migrated outwards.  The length of
time used to measure properties of encounters
 is also reported in Table \ref{tab:tab1} in orbital periods of the innermost planet. 
Final positions of the planets $a_{1,f}$ and $a_{2,f}$ and the total number of orbit crossing planetesimals, $N_c$,
are also listed in Table \ref{tab:tab1}.
 The change in total energy due to numerical integration errors
at the end of the integrations was
$|\Delta E/E_0| \lesssim 10^{-4}$ where $E_0$ is the initial energy (see \citealt{moore11} for a
discussion about the accuracy of this integrator).
The simulations were run using NVIDIA graphics cards from the GT200 architecture (the Quadro FX5800, or
the GTX 285) or using Fermi Class Tesla C2050 cards from the GF100 architecture.  All of these are capable
of computing in double precision. 

\begin{table*}
\vbox to135mm{\vfil 
\parbox{4in}{\caption{\large Integrations  \label{tab:tab1}}}
\begin{tabular}{@{}lrccccccc}
\hline
$M_1$     & $M_2$                  & $P$     &$a_{1,f}$&$a_{2,f}$&$N_c$ & $\log_{10}P_1$ & $\log_{10} P_2$\\
(1)            & (2)                        & (3)       & (4)         &  (5)         &     (6) & (7)      & (8)       \\
\hline
$10^{-3}$ &$5 \times 10^{-5}$ &  1200  &  0.99    & 2.3           & 5345  &  -4.43  & -1.20 \\  
$10^{-3}$ &$7 \times 10^{-5}$ &  1500  &  0.98    & 2.3           & 6042  & -3.58   & -1.26 \\
$10^{-3}$ &$             10^{-4}$ &   1910 &   0.97    & 2.3          & 6728  &  -3.83 &  -1.31 \\   
$10^{-3}$ &$3 \times 10^{-4}$ &  1910 &   0.93    & 1.9          & 5234   &  -3.94 &  -1.53  \\ 
$10^{-3}$ &$4 \times 10^{-4}$ &  1910 &   0.92    & 1.8          & 4700   &  -3.53  & -1.58 \\
$10^{-3}$ &$6 \times 10^{-4}$ &  1910 &   0.90    & 1.7          &  4494  &  -3.57  & -1.66 \\
$10^{-3}$ &$7 \times 10^{-4}$ &  1910  &  0.89    & 1.7          & 4738   &  -3.47  & -1.72 \\
$10^{-3}$ &$8 \times 10^{-4}$ &  1910  &  0.88    & 1.7          & 5108   & -3.33   & -1.76 \\
$10^{-3}$ &$              10^{-3}$ & 1910 &   0.88    & 1.6           & 4617   & -2.87  & -1.84   \\   
$10^{-4}$ &$2 \times 10^{-5}$ & 1400  & 1.0        & 2.2          & 4732 & -3.76  & -1.36 \\
$10^{-4}$ &$3 \times 10^{-5}$ & 1500  & 1.0        & 2.2          & 5310  & -3.99  & -1.32 \\  
$10^{-4}$ &$5 \times 10^{-5}$ &  1910 & 1.0        & 2.2          &  3628 & -4.12  & -1.22  \\ 
$10^{-4}$ &$              10^{-4}$ &  1910 & 1.0       & 1.8          & 3316   & -3.78  & -1.34  \\ 
%
$8 \times 10^{-4}$ &$8 \times 10^{-5}$ & 1500  & 0.98 & 2.3 &  6031 & -3.80 & -1.27 \\      
$2 \times 10^{-3}$ &$2 \times 10^{-4}$ & 1910  & 0.94 & 2.2 &  6174 & -3.54 &  -1.47 \\ 
$3 \times 10^{-4}$ &$9 \times 10^{-5}$ & 1910  & 0.98 & 2.2 &  5625 & -4.06 &  -1.30 \\
$6 \times 10^{-4}$ &$9 \times 10^{-5}$ & 1910  & 0.98 & 2.3 &  6182 & -3.83  & -1.30 \\
\hline
\end{tabular}
{\\ \parbox{3.9truein}{
This table lists properties of each numerical integration.
First and second columns give the masses of the planets in units of the stellar mass;  $M_1$, $M_2$. 
The time $P$ in orbital periods of the innermost planet (based on its initial orbital period) used to compute statistical properties of encounters are listed in column 3.
The semi-major axes of the planets after $P$ orbital periods are listed as $a_{1,f}, a_{2,f}$ (columns 4,5). The total
number of unique planet orbit crossing planetesimals identified is listed in column 6.
Column 7 shows the log of the probability that the inner planet would capture the secondary of a planetesimal binary due to a binary exchange reaction for a binary that disrupts at a normalized
tidal disruption radius $R_{td}=0.1$ (as discussed in section 3).  This probability is given for planetesimal binary
with a primary of radius  $s_1 = 100$ km but can be scaled to other planetesimal binaries using
factors given in equation \ref{eqn:pcap}.   
The normalized disruption radius, $R_{td}$ is defined in equation \ref{eqn:Rdisrupt}.  
Column 8 lists the log of a similar probability but for the outer planet. The probabilities of capture
have been estimated from the encounter distributions in the simulations and using equation (\ref{eqn:pcap}).}\\ 
}
 \vfil}
\end{table*}

\begin{figure}
\includegraphics[width=9cm]{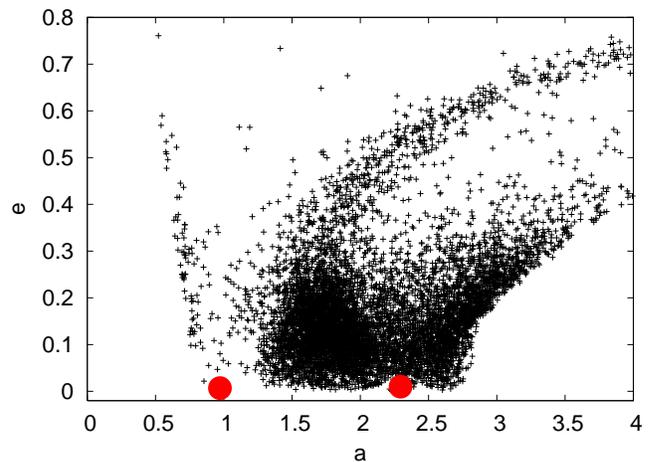}
\caption{Semi-major axis and eccentricities of particles at the end of the integration with planet masses 
$M_1 = 10^{-3}$
and $M_2=10^{-4}$.  Solid circles
are the planets.  Scattering surfaces are seen for both planets at eccentricities setting pericenters and apocenters
near the planets.  The outer planet has migrated more than half way through the planetesimal disk.
}
\label{fig:ae}
\end{figure}

Planetesimals originating in the outer disk are first scattered by the outer planet.  Afterwards they
 can cross the orbits of both planets.  
The interaction between planetesimals and planets allows the outer planet to migrate outwards into
the planetesimal disk via the exchange of angular momentum between planets
and planetesimals \citep{fip84}.
While planetesimals are scattered by planets they can have pericenter near that of a planet's semi-major 
axis.  On a plot of semi-major axis versus eccentricity these can be seen as scattering surfaces associated
with each planet, as seen in previous integrations of planet migration (e.g., \citealt{kirsh09}).
To illustrate this, Figure \ref{fig:ae} 
shows the semi-major axis and eccentricity distribution at the end
of the integration with $M_1=10^{-3}$ and $M_2=10^{-4}$.   
At this time approximately half of the planetesimals have experienced encounters
with a planet.  The other integrations look similar but with different fractions of particles in each scattering surface
and differing numbers of particles in between the planets in orbits that cross the orbits of the two
planets.

\subsection{Trajectories of planet crossing particles}

\begin{figure}
\includegraphics[width=9cm]{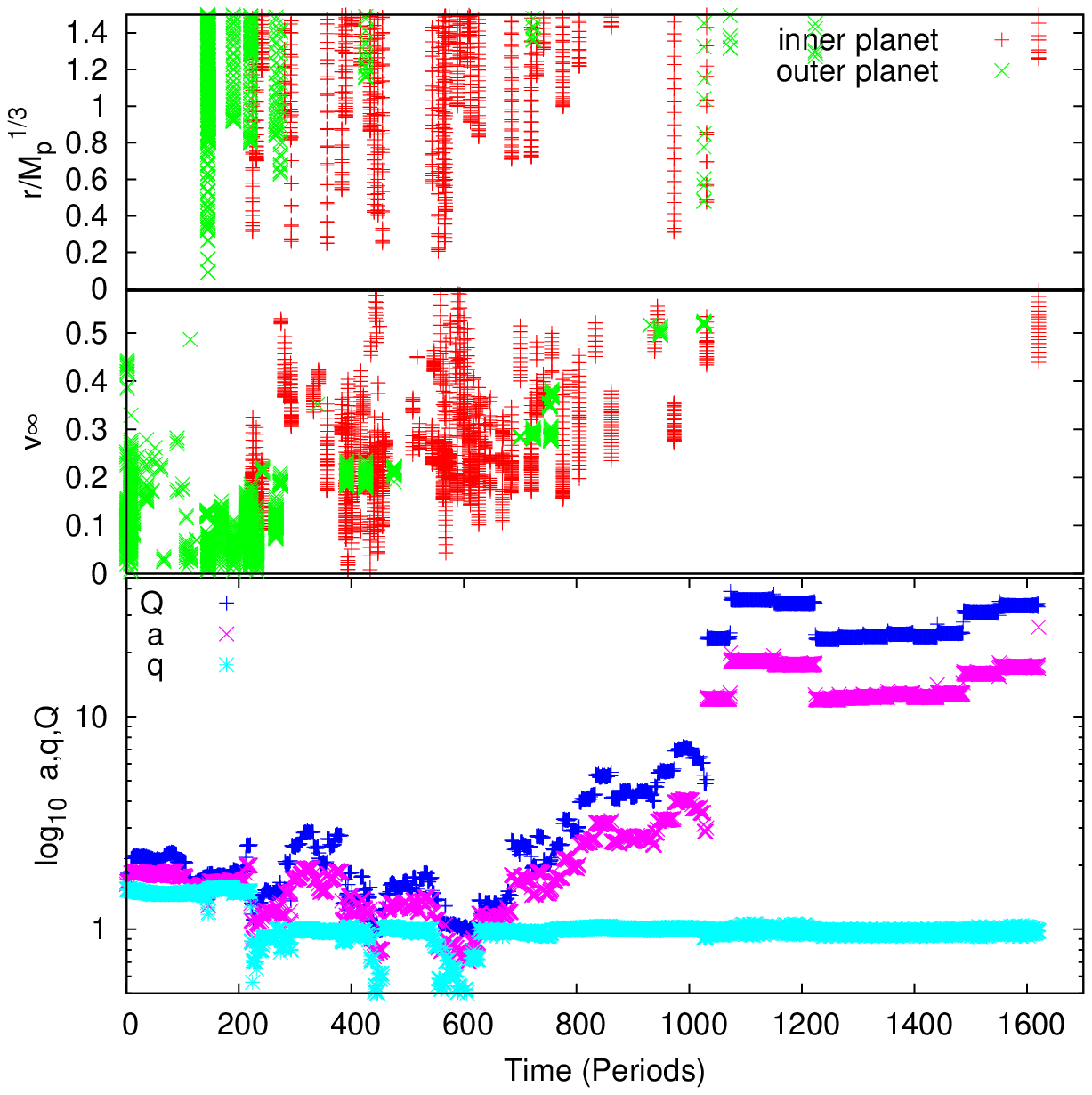}
\caption{Evolution of a planetary orbit crossing particle.  
Bottom: We show the log of the 
semi-major axis (purple crosses), pericenter (light blue stars) and apocenter (dark blue plus signs)
as a function of time for a particle that was eventually ejected in the integration with planet
masses $M_1=10^{-3}$ and
$M_2 = 3 \times 10^{-4}$.
Changes in particle semi-major axes and eccentricity are caused by close encounters with planets.
Middle: The velocity with respect to each planet during close encounters.  Green points show
velocities for encounters with the outer planet and red points show those for the inner planet. We note that
early encounters with the outermost planet tend to have the lowest relative velocities.
Top: the distance between
the object and planet divided by the planet mass to the third power during close encounters.   
Green points show
distances to the outer planet and red points show those to the inner planet.
The closest approach at $t \sim 200$ occurs early in the simulation and with the outer planet.
This approach also is slow and so could allow the secondary of a binary planetesimal to be captured into orbit
about the outer planet during binary disruption. 
The order of
the encounters can be important as an early fast close encounter can disrupt a binary making
subsequent irregular satellite capture impossible even if the individual planetesimals later on
experience slower encounters.
\label{fig:traj}}
\end{figure}

Because we have logged all particle/planet encounters during the integrations we can examine 
trajectories of individual particles.
In Figure \ref{fig:traj} we show an example of the trajectory of a disk particle from the simulation 
with $M_1=10^{-3}$ and $M_2 = 3\times 10^{-4}$ that was ejected
near the end of the simulation. We identify ejected
particles based on their energies (computed with respect to the central star); 
those with positive energy have been ejected from the system.
In this figure we show the particle's semi-major axis, $a$, pericenter, $q$, and apocenter, $Q$, 
as a function of time.
We also show simultaneously distance between particle and each planet and the velocity $V_\infty$ computed
with equation (\ref{eqn:vinf}) as described above.  Encounters are only logged when a particle passes within
a Hill radius of a planet so $V_\infty$ and $r$ are only plotted during encounters.

When the particle first crosses the orbit of the outer planet it is in a low eccentricity orbit.  The velocity of
the particle with respect to the planet is not large and so encounters can be gravitationally focused.
We see in Figure \ref{fig:traj} that a slow and close approach to the outer planet 
happens before close approaches to the inner planet.  The eccentricity of the particle
must increase before its orbit can cross those of both planets. The velocity difference between particle
and planet is larger when there is a large difference between particle and planet semi-major axis and eccentricity. 
The inner planet is larger than the outer one in this integration and so the inner planet exchanges 
more energy with the particle. 
The trajectory shows a series of
distant approaches to the inner planet that slowly increase the energy of the particle.  
Ejection is not due to a single close encounter with the inner planet.   
If this particle were a binary planetesimal then it would likely disrupt during one of the slow 
and close approaches with the
outer planet just after it becomes orbit
crossing rather than later on when its semi-major axis is high and the relative velocity
between planets and particle higher.   The trajectory shown in Figure \ref{fig:traj} 
also illustrates that the order of close encounters
with the planets is important.   A fast but close encounter could disrupt a binary without allowing a capture to 
take place, hence both the order of encounters and the velocities of disrupting encounters 
must be taken into account to predict the probability of satellite capture.
In the next subsection we compute histograms of disruption locations and velocities during disruption
for each planet.

\subsection{Histograms of disruption locations}

A planetesimal binary composed of two masses $m_1,m_2$ and  separation $a_B$ in the vicinity of a planet
with mass $M_p$ disrupts at a tidal disruption radius
\begin{equation}
r_{td} \sim a_B \left({3M_p \over m_1 + m_2}\right)^{1/3} 
\label{eqn:rtd}
\end{equation}
(e.g., \citealt{agnor06}).
A planetesimal binary that becomes planet orbit crossing will be disrupted the first time it comes
within a tidal disruption radius of a planet.
Previous studies have measured distributions of closest approaches in different contexts \citep{levison97,vok08} but
have not measured the distribution of disruption radii that depend on both planet mass and the order of
close approaches for the idealized migrating system studied here.

A binary disrupts if it passes within its tidal disruption radius of a planet (equation \ref{eqn:rtd}).  The ratio,
 $r_{td}/M_p^{1/3}$,   only depends on binary properties (its separation and total mass).  
 For each encounter that approaches within a radius $r_{td}$ of a planet with mass $M_p$ 
 we compute the normalized disruption distance
 \begin{equation}
 R_{td} = {r_{td} \over a_{1,init}}\left({ M_* \over M_p}\right)^{1/3}
 \label{eqn:Rdisrupt}
 \end{equation}
 where $a_{1,init}$ is the initial semi-major axis of the inner planet.  
 A binary that disrupts at $R_{td}$ has $R_{td} = {a_B \over a_{1,init}} \left({ 3M_* \over m_1+ m_2}\right)^{1/ 3}$
 only dependent on binary properties.  This is why we will measure encounter distances in
 units of $R_{td}$.
  
 We now consider the range of possible values for $R_{td}$.
 The planet's Hill radius is
\begin{equation}
r_{Hp} \equiv a_p \left( { M_p \over 3M_*}\right)^{1/3}.
\end{equation}
The normalized disruption distance, $R_{td}$, differs from $r_{td}/r_{Hp}$ by
a ratio of planet semi-major axes.  However since we expect the ratio of planet semi-major axes
is of order unity the maximum value of $R_{td}$ is approximately 1.
The minimum value of $R_{td}$ occurs when a binary has separation, $a_B$, approximately equivalent to
the radius of its more massive body. This gives a limit
\begin{equation}
R_{td} \lesssim 10^{-3} \left( {\rho \over 1 ~ {\rm g~cm}^{-3}}\right)^{-{1\over 3}}
 \left( {M_* \over M_\odot}\right)^{1\over 3}
  \left( {a_{1,init}\over 10~{\rm AU}}\right)^{-1}
\end{equation}
where $\rho$ is the density of the planetesimals.  The range of allowed values for $R_{td}$ sets the range covered
by the $x$-axes of our subsequent figures.

\begin{figure}
\includegraphics[width=9cm]{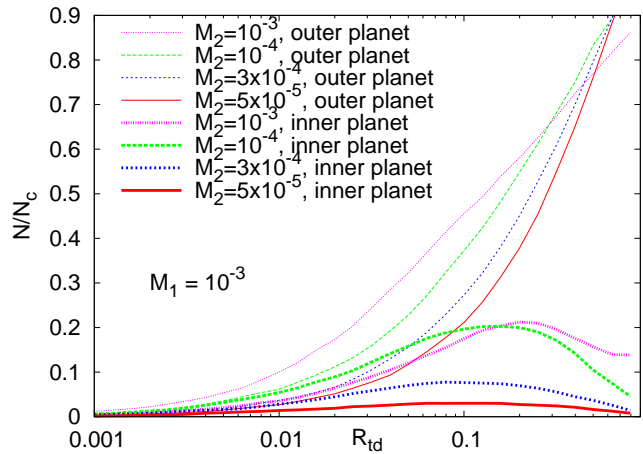}
\caption{The fraction of orbit crossing particles ($y$-axis) that pass within a disruption distance of 
$R_{td}$  ($x$-axis) as defined by equation (\ref{eqn:Rdisrupt})
of each planet for four simulations each containing two planets.  The properties of the binary
(mass and separation) determine the radius at which it would disrupt or $R_{td}$.
Four simulations are shown here and all have inner planet with mass $M_1 = 10^{-3}$.
The outer planet mass is $M_2 = 0.5, 1, 3 $ or $10 \times 10^{-4}$ depending upon the simulation.
The top four and thin curves show fractions of particles that approach the outer planets
in simulations (from top to bottom) for simulations with $M_2 = 0.5, 1, 3 $ and $10 \times 10^{-4}$.
The bottom four and thicker curves show fractions for the inner planet for the same simulations.
Simulations are labelled by the mass of the outer planet. 
At each tidal disruption distance, each particle is only counted once and for the planet at which it first approaches
within the disruption distance.  Weakly bound binaries are preferentially disrupted by the outer planet (high
curves on the right hand side of the plot).
The difference between the fraction disrupted by the outer planet and disrupted by the inner planet increases
as a function of the ratio of inner to outer planet mass. 
}
\label{fig:qhist}
\end{figure}

For each planet, we count the number of particles
that pass within a given disruption radius. 
At each disruption distance each orbit crossing particle is only counted once.  A particle
is counted at a disruption distance for the planet that it {\it first} approaches within this distance.
The result is two histograms (one for each planet) that are functions of $R_{td}$ and that can be used
to predict the disruption radius for binaries of a given mass and separation and to predict the planet 
that is responsible for the disruption.  
The resulting histograms for simulations with $M_1 = 10^{-3}$ and
$M_2 =  \{ 0.5, 1, 3,$ and  $10\}  \times 10^{-4}$ are shown in Figure \ref{fig:qhist}.  In these simulations 
about half of the simulated planetesimals have passed within the Hill radius of a planet. 
The histograms are normalized by the number of planet orbit crossing particles, $N_c$, with this quantity listed
for each simulation in Table (\ref{tab:tab1}).   

Because each particle is only counted one (at each disruption radius) we can determine which planet 
causes the disruption for each type of binary.
Wide and low mass binaries are disrupted at large disruption radii whereas tight and massive binaries disrupted
at smaller disruption radii.
In Figure \ref{fig:qhist} we see that wide binaries are most likely to tidally disrupt near the outer planet.  For
simulations with similar planet masses, 
tightly bound binaries are more likely to be disrupted by the inner planet.
More binaries are disrupted at each disruption radius when the planets are more massive.  
The difference between the fraction disrupted by the outer planet and that disrupted by the inner planet is 
largest when the planet mass ratio is large. 
We expect that
an outer planet would have a much richer irregular satellite population than an inner but more massive planet.

\subsection{Velocity Distributions}

\begin{figure}
\includegraphics[width=9cm]{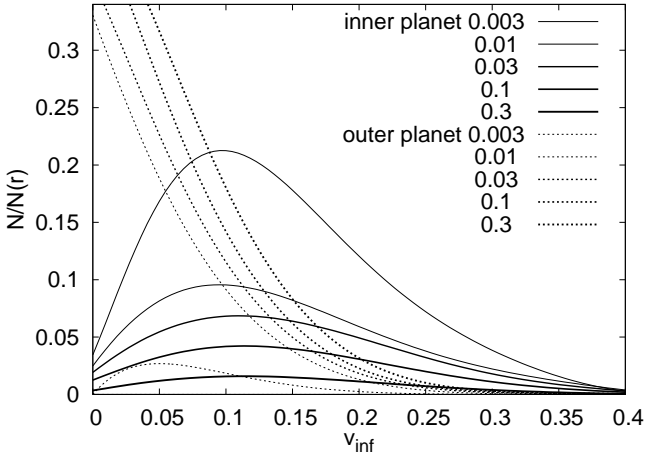}
\caption{Velocity histograms for the simulation with planets $M_1=10^{-3}$ and $M_2=10^{-4}$ 
that is also displayed in Figure \ref{fig:qhist}.  
For 5 different disruption radii ($R_{td}$) we plot the distributions 
of incoming velocities, $V_\infty$,
for disrupting close encounters for each planet.   The disruption radii are labelled in the key on the upper right.
Thick lines correspond to the velocity distributions for the inner planet and thin lines
for those of the outer planet.  
Velocities are in units of the inner planet's initial circular velocity.
The top dotted curve is that for the inner planet and $R_{td}=0.3$ and the bottom solid curve
is that for the outer planet and $R_{td}=0.3$.
Encounters with the outer planet are more likely to have lower encounter velocities than the inner planet, consequently 
the outer planet is more likely to capture a secondary during tidal disruption of a planetesimal binary.
}
\label{fig:vhist}
\end{figure}

During tidal disruption the probability that the secondary of the planetesimal binary is left bound to the planet
depends on the incoming velocity of the binary \citep{agnor06}.   The secondary is more likely to be
captured into orbit if the incoming velocity is low.
For 5 different disruption radii we plot in Figure \ref{fig:vhist} velocity distributions of the disrupting encounters
for each of the two planets for the simulation with $M_1 = 10^{-3}$ and $M_2 = 10^{-4}$ (which is
also displayed as one of the pairs  of curves  in Figure \ref{fig:qhist}).  
The histograms give the fraction of particles with each $V_\infty$ value 
(with $V_\infty$ computed using equation \ref{eqn:vinf}).
These plots have been normalized so that at each disruption radius the sum 
of the integrated velocity distributions for the two planets is 1.      
The curves have been smoothed and particles
with computed negative $E_p$ neglected from the histograms.  
During encounters some particles have computed negative $E_p$ and appear to be bound when they are not
because the motion of the planet with respect to the central star has been neglected in the computation of $E_p$. 

We see in Figure \ref{fig:vhist} that
the velocity distributions differ for each planet with the mean of the distributions significantly 
higher for the inner planet than the outer planet.    
Low velocity encounters are much more likely with the outer planet than the inner planet.
Tight binaries can  be disrupted by the inner planet but have sufficiently high velocities that the capture
of the secondary into orbit around the planet is unlikely.    The curves associated with the outer planet
are higher at zero velocity than those for the inner planet implying that orbit crossing planetesimal
binaries are more likely to leave a bound object around the outer planet even if they are more likely to be
disrupted by the inner planet.  In the next section we estimate the probability of irregular satellite capture based on
the low velocity region of these distributions for our different numerical integrations.

As seen from Table \ref{tab:tab1} for most of our integrations the outer planet migrates a substantial 
way through the planetesimal disk.    By dividing integrations into different time intervals and comparing
the distributions based on encounters in these intervals  we have checked that the distributions based on the encounters  (as shown in Figures \ref{fig:qhist}, \ref{fig:vhist} and subsequent figures)
are not strongly dependent upon time.
At the beginning of the integrations the outer planet is near 
the inner edge of the planetesimal disk.   
We have found that
as long as this edge is sufficiently close to the outer planet to allow migration to take place during the integration, 
 the encounter distributions are insensitive to the disk edge location.
We attempted to determine if the distributions were sensitive to disk mass by increasing or decreasing
the mass of the planetesimals.  However we cannot significantly increase the number of disk particles
without increasing the length of computation time.  If we significantly decrease the disk
or planet masses then planet migration can be extremely slow.   Each simulation took a few days
to run, consequently we were effectively limited to a narrow region of parameter space in both 
planet and disk masses.
We compromised by fixing the disk mass and then choosing only simulations in which the outer planet
migrated a significant distance through the disk.    

\section{Exchange reactions}

We now consider the properties of a planetesimal binary distribution that could lead to capture of an irregular
satellite population in a migrating planetary system.   We consider a planetesimal binary composed of two masses $m_1,m_2$ and  separation, $a_B$, in the vicinity of a planet.  During tidal disruption in a planet's tidal 
gravitational field,
the change in velocity of the smaller body, $m_2$, due to the orbital speed about the binary's center of mass
\begin{equation}
\Delta v_2 \approx {m_1 \over m_1 + m_2} \sqrt{ G (m_1 + m_2) \over  a_B} = {m_1 \over m_1 + m_2} v_B
\label{eqn:deltav2}
\end{equation}
(following \citealt{agnor06})
where the binary internal velocity
\begin{equation}
v_B \equiv \sqrt{G(m_1 + m_2) \over a_B}. 
\end{equation}
To enable capture, the difference in energy due to the velocity change must exceed the initial energy
of the object or $m_2 V_\infty^2/2$ where $V_\infty$ approximates the velocity (with respect to the planet)
before the planet enters the
planet's Hill sphere (equation \ref{eqn:vinf}).   
This condition is easiest to satisfy for the secondary rather than the primary (e.g., \citealt{gould03}) hence
that is the situation we focus on here.

We define the Hill radius of the binary at its birth semi-major axis $a_{birth}$,
\begin{equation}
r_{HB} \equiv a_{birth} \left( {m_1 + m_2 \over 3 M_* }\right)^{1/3}.
\end{equation}
The ratio of the tidal disruption radius to the planet's Hill radius
\begin{equation}
{r_{td} \over r_{Hp} } \sim \left({a_B \over r_{HB}} \right) \left({a_{birth} \over a_p}\right) 3^{1/3} 
\label{eqn:rtdH}
\end{equation}
where $a_p$ is the semi-major axis of the planet.
A binary cannot have a separation larger than its own Hill radius.
The above equation implies that the disruption location in units of a planet's Hill radius
primarily depends on the binary separation in units of its own Hill radius.
As mentioned in section 2.3, $R_{td}$ differs from $r_{td}/r_{Hp}$ by a factor of planet radii.
Thus the normalized disruption radius $R_{td}$ given as the $x$-axis in Figure \ref{fig:qhist} 
is approximately the ratio of the binary's Hill radius to that of the planet.
Transneptunian binaries tend to have separations of order $a_B/ r_{HB} \sim 0.01$  (range 0.002-0.03) 
\citep{grundy11} and so would disrupt at radius (from the planet) of about 0.01 a giant planet's  Hill radius
or with $R_{td}$ of order 0.01.


As the capture probability also depends on the binary's internal circular velocity, $v_B$, 
(see equation \ref{eqn:deltav2})
it is useful to consider the maximum and minimum values
for this  velocity.
The binary must be bound so that $a_B < r_{HB}$ leading to a minimum binary internal velocity
\begin{eqnarray}
 \label{eqn:vbmin}
{v_B \over v_{KB} } &\gtrsim & 3^{1 \over 6} \left(m_1 + m_2 \over M_* \right)^{1 \over 3} \\
&\gtrsim& 10^{-4} \left({s_1 \over 100 {\rm km}} \right) \left({ \rho \over 1 {\rm g~cm}^{-3}}\right)^{1\over 3} 
\left({M_* \over M_\odot} \right)^{-{1 \over 3}} (1+ q_b)^{1\over 3} , \nonumber
\nonumber
\end{eqnarray}
where $v_{KB}$ is the Keplerian circular velocity at the binary's semi-major axis 
(around the star) and $q_b\equiv m_2/m_1$
is the binary mass ratio.
The above is equivalent to requiring $v_B > v_{HB}$ where the Hill velocity $v_{HB} = \Omega_K r_{HB}$ 
and $\Omega_K$ is the Keplerian angular rotation rate (or mean motion around the star) 
at the binary's semi-major axis.
When the two planetesimals are nearly touching the binary has   a maximum internal velocity of
\begin{eqnarray}
\label{eqn:vbmax}
{v_B \over v_{KB} } &\lesssim& \sqrt{\rho s_1^2 a \over M_* } \\
&\lesssim    &
 0.003 \left({\rho \over 1 {\rm g~cm}^{-3}}\right)^{1\over 2} 
\left({ M_* \over M_\odot}\right)^{-{1 \over 2}}   \nonumber \\
&& \times \left({ a\over 10 {\rm AU}}\right)^{1 \over 2} 
   \left({ s_1 \over 100 {\rm km}}\right). \nonumber
\end{eqnarray}
The range of binary internal velocities is restricted by these two limits and
is not large.  The maximum is larger when the
binary in the outskirts  of a planetary system as is true for transneptunian binaries.

\subsection{Estimating the Probability of Irregular Satellite Capture}

In the above section we showed that a binary planetesimal disrupts at a distance from a planet in Hill
radii that is primarily dependent on the binary separation in units of its own Hill radius (equation \ref{eqn:rtdH}).
A planetesimal binary could be tidally disrupted but both objects will remain unbound unless the disruption
event removes sufficient energy that one of the objects becomes bound to the planet.  
We now consider the probability
that the secondary (from the binary) is captured as a function of the binary's incoming velocity, $V_\infty$,
with respect to the planet.  We will derive a limit on $V_\infty$ that allows a capture to take place. Using this
limit we will use our simulations (in which we measure $V_\infty$ for close approaches) to estimate
the probability that the secondary of a binary planetesimal can be captured into planetary orbit.
  
The energy per unit mass of a binary approaching a planet
\begin{equation}
E_0 \approx {V_\infty^2 \over 2},
\end{equation} 
where we have only taken into account the gravitational field of the planet and neglected that from the star.
At the tidal radius, $r_{td}$, the energy per unit mass can be written
\begin{equation}
E_0 = {{\bf v}(r_{td})^2 \over 2} - {GM_p \over r_{td}},
\end{equation}
where ${\bf v}(r_{td})$ is the velocity of the orbit at disruption.
For the secondary, $m_2$, of the binary to become bound to the planet, the change in velocity 
during disruption must remove sufficient energy that the energy, $E$, becomes negative.
The change in energy due to a velocity kick at $r_{td}$ caused by disruption
\begin{equation}
\Delta E = {({\bf v}(r_{td}) + \Delta {\bf v}_2)^2  \over 2} - {v (r_{td})^2 \over 2}= {\bf v}(r_{td}) \cdot \Delta {\bf v}_2 + {(\Delta v_2)^2 \over 2} 
\end{equation}
with $|\Delta v_2|$ given by equation (\ref{eqn:deltav2}) and $\Delta E \gtrsim V_\infty^2/2$ required for capture.
Let the angle between the two vectors ${\bf v}(r_{td})$ and $\Delta {\bf v}_2$ be $\theta_v$.
Using the previous equation and this angle, for capture we require that
\begin{equation}
 |\Delta v_2|| v(r_{td})| \cos \theta_v + {(\Delta v_2)^2 \over 2} \gtrsim {V_\infty^2 \over 2}.
 \label{eqn:determs}
\end{equation}

Because the incoming binary is not bound to the planet, the orbital velocity at the moment of tidal disruption, $v(r_{td})$,
 must always exceed the escape velocity at $r_{td}$.
 The escape velocity at the tidal disruption radius is
 \begin{equation}
v_{escape} (r_{td}) = \sqrt{ 2 G M_p \over r_{td} }
= v_B 2^{1 \over 2} \left( { M_p \over m_1 + m_2}\right)^{1 \over 3}. \label{eqn:ves}
 \end{equation}
We can consider two limits. When the orbit is nearly parabolic, the orbit speed
at disruption $v (r_{td}) $ is 
approximately equal to the escape velocity $v (r_{td}) = \sqrt{2 G M_p \over r_{td}}$.
If the orbit is highly hyperbolic then $v (r_{td})$ is approximately equal to $V_\infty$. 
The gravitational focusing factor
\begin{equation}
A_f = \sqrt{ 2G M_p \over r_{td} V_\infty^2 } = {v_{escape} (r_{td}) \over V_\infty} = {v_B \over V_\infty}2^{1 \over 2} \left( { M_p \over m_1 + m_2}\right)^{1 \over 3}
\label{eqn:focus}
 \end{equation}
 determines the regime; when this factor is low ($A_f <1$) the orbit is highly hyperbolic.

 It is useful to compare the escape velocity at the tidal disruption radius
 to the velocity kick $\Delta v_2$.
 Using equations (\ref{eqn:ves}) and
 (\ref{eqn:deltav2}) the ratio
 \begin{equation}
 {v_{escape}(r_{td}) \over \Delta v_2 }  = 2^{1\over 2} 3^{-{1 \over 6}} \left( { M_p \over m_1 + m_2}\right)^{1 \over 3}
\left({ m_1+m_2 \over m_1 } \right).
 \end{equation}
 Since we expect nearly equal mass binaries the above ratio should exceed 1.  As $v_{escape}(r_{td})$
 is a lower bound on the orbit velocity at disruption (and $v_{escape}(r_{td} > \Delta v_2 $) we find that
 the first term in equation (\ref{eqn:determs}) should dominate the second term.  
 Dropping the second term in equation
 (\ref{eqn:determs})  we can approximate the condition
 for capture as
 \begin{equation}
  \cos \theta_v  \gtrsim {V_\infty^2 \over 2 v(r_{td}) \Delta v_2} . 
  \label{eqn:decond}
 \end{equation}
 
 Assuming a random distribution of binary orientations,
we can integrate the fraction of the spherical angle that satisfies the above condition (equation \ref{eqn:decond}).
We can consider a critical angle, $\theta_c$, that is the value of $\theta_v$ that
gives equality in equation (\ref{eqn:decond}).
Integrating over solid angle (assuming that all binary orientations equally probable) 
the probability of capture, $p_c$,  depends on this critical angle, $\theta_c$, with
\begin{equation}
p_c \sim {\sin \theta_c \over 2}
\end{equation}

Using equation (\ref{eqn:decond}) 
the strongly hyperbolic orbit ($A_f <1$ and $v(r_{td}) \sim V_\infty$) gives  for capture
\begin{equation}
\cos \theta_c \sim {V_\infty \over 2 \Delta v_2}  \sim {V_\infty \over v_B} { m_1 + m_2 \over 2 m_1}.
\label{eqn:hyp}
\end{equation}
The nearly parabolic orbit ($A_f >1$ and $v(r_{td} )\sim v_{escape} (r_{td})$) gives
\begin{eqnarray}
\cos \theta_c &\sim& {V_\infty^2 \over 2 v_{escape} (r_{td})\Delta v_2 } \nonumber \\
&\sim& \left({V_\infty \over v_B}\right)^2  \left( { m_1 + m_2 \over 2^{3\over 2} m_1}\right)
\left({ m_1 + m_2 \over M_p} \right)^{1\over 3}. 
\label{eqn:para}
\end{eqnarray}
Note that
 $\cos \theta_c$ is at maximum 1 if the right hand side of either the previous two expressions 
 exceeds 1.  Both of these expressions primarily depend on the ratio $V_\infty/v_B$. Capture is only
 likely if $\cos \theta_c$ is less than 1 and so can only take place if $V_\infty/v_B$ is small.
 Since $V_\infty/v_B$ must be small then $A_f \propto v_B/V_\infty$ must be large (see equation
 \ref{eqn:focus}) implying that the encounter must be nearly parabolic for capture to take place.
 This means we can use equation (\ref{eqn:para}) and ignore equation (\ref{eqn:hyp}).
 %
 Furthermore the probability drops from 0.5 to 0 over 
 a very small range of values for $V_\infty$ and mass ratio $M_p/(m_1+m_2)$.   
 We can estimate the condition for capture based on
 setting the equation (\ref{eqn:para}) to 1 and determining how $V_\infty$ depends
the properties of the binary at this this dividing line. Setting $\cos \theta_c = 1$ and inverting equation
 (\ref{eqn:para}) we find that
 capture of the secondary during tidal disruption is likely (and has a probability of about 1/2) when
 \begin{equation}
 {V_\infty \over v_B }\lesssim \left( { M_p \over m_1 + m_2}\right)^{1\over 6} (1 + q_b)^{-{1\over 2}} 2^{3\over 4}
 \label{eqn:vib}
 \end{equation}
 and is nearly zero otherwise. Here the binary mass ratio $q_b \equiv m_2/m_1$.
 While tidal disruption only depends on radius (from the planet) in units of Hill radii, we see here
 that the probability of capture
 also depends on $V_\infty /v_B$.  As more compact and massive binaries have larger orbital velocities
 they are more likely to leave the secondary as an irregular satellite during a close approach (as previously discussed by \citealt{agnor06,philpott10}).  Here we have a quantitative, though approximate
 condition on $V_\infty/v_B$ allowing capture which we can use to estimate capture probabilities from
  the encounter velocity distributions measured in our simulations.
 
In our integrations we measure velocities in units of the Keplerian rotation velocity (about the star) of the inner planet ($v_{K1}$), hence it is useful to write
our condition for capture (equation \ref{eqn:vib})  in terms of $V_\infty/v_{K1}$ or 
\begin{equation}
{V_\infty \over v_{K1}} \lesssim \left( { M_p \over m_1 + m_2}\right)^{1\over 6}
\left({ (m_1 + m_2) \over M_*}{a_{1,init} \over  a_B} \right)^{1\over 2} (1 + q_b)^{-{1 \over 2}} 2^{3\over 4}
\label{eqn:vinfvk}
\end{equation}
Rather than describe binaries as a function of mass, $m_1 + m_2$, and separation, $a_B$,
we can describe a binary in terms of its mass and the normalized tidal disruption ratio $R_{td}$ 
(defined in equation \ref{eqn:Rdisrupt}).
The condition for capture in equation (\ref{eqn:vib}) or (\ref{eqn:vinfvk}) becomes
\begin{eqnarray}
\label{eqn:vinfvk2}
{V_\infty \over v_{K1}} &\lesssim& \left( { a_{1,init} \over r_{td} } {M_p^{1/3} \over M_*^{1/3} }\right)^{{1 \over 2}}
	\left({ M_p\over M_* }  \right)^{1\over 6}
	\left({ m_1+ m_2 \over M_* }  \right)^{1\over 6} \nonumber \\
	&& \times  (1+q_b)^{-{1 \over 2}} 2^{3 \over 4} 3^{1 \over 6} \\
	\label{eqn:vinfvk2b}
&\sim & 0.02 \left( { R_{td} \over 0.1} \right)^{-{1 \over 2}} 
	\left({ M_p/M_* \over 10^{-3} }\right)^{1\over 6}
	\left({\rho \over 1 ~{\rm g~cm}^{-3}} \right)^{1\over 6} \nonumber \\
&& \times \left({ s_1 \over 100 ~{\rm km}} \right)^{1 \over 2} 
\left({ M_* \over M_\odot }\right)^{-{1 \over 3}}(1 + q_b)^{-{1 \over 3}}. 
\end{eqnarray}

\begin{figure}
\includegraphics[width=9cm]{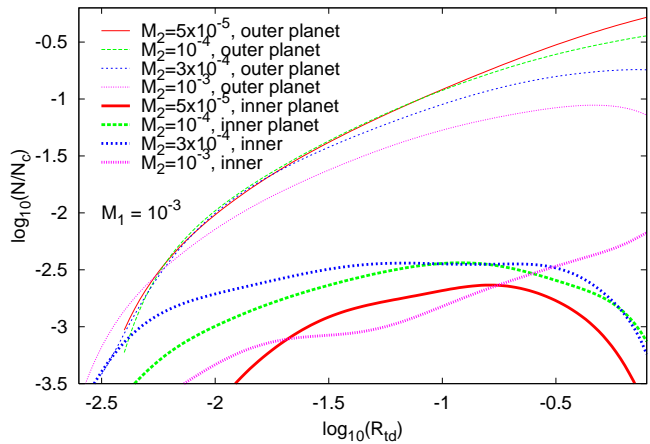}
\caption{Fraction of integrated particles with incoming velocity $V_\infty$  between 0 and 0.05 
(in units of the inner planet's 
Keplerian velocity) for the integrations shown in Figure \ref{fig:qhist} as a function of different normalized binary disruption radii, $R_{td}$. 
The curves have been normalized by the number of orbit crossing particles in the simulation, $N_c$.   
Thick lines correspond to the encounters with inner planets and thin lines to those with the outer planets.
In all cases slow encounters are more likely with the outer planet and so this planet is more likely than the inner planet to capture satellites through binary planetesimal exchange reactions. 
}
\label{fig:v0hist}
\end{figure}

We have estimated that the probability of capture is about 1/2 when the incoming velocity, $V_\infty$,
is less than the expression we give in equation (\ref{eqn:vinfvk2}).   
Our numerically measured distributions of $V_\infty$ can now be used to estimate the probability that a distribution
of planetesimals will leave behind some captured secondaries.
 Because only the lowest velocity encounters are relevant we can take the velocity distributions measured
 from the simulated encounters,  shown in Figure \ref{fig:vhist},
 and focus on only the number of encounters with incoming velocity near zero.  
 Figure \ref{fig:v0hist} plots
 the fraction of encounters $N(R_{td}) /N_c$ passing within a disruption distance that have 
 $V_\infty$ within 0 to 0.05 (the value
 in units of the innermost planet's Keplerian velocity) for the same simulations shown in Figure \ref{fig:qhist}.  
 
 The fraction of encounters with velocities below that given by equation (\ref{eqn:vinfvk2})
 can be estimated by multiplying our numerically measured fraction $N(R_{td})/N_c$ 
 integrated over values of $V_\infty \in [0,0.05]$ by the factor
 \begin{equation}
 f = {0.02\over 0.05} \left({R_{td} \over 0.1}\right)^{-{1\over 2}} \left({M_p/M_* \over 10^{-3}}\right)^{1\over 6}.
 \label{eqn:fac}
 \end{equation}
 Here the factor of 0.02,  and $R_{td}^{-1/2} M_p^{1/6}$ are those
 in equation (\ref{eqn:vinfvk2b}). We convert this normalized fraction 
 to a probability of irregular satellite capture using an additional factor of 1/2 since
 the probability of capture is at most 1/2.  
 Thus the probability of a planetesimal capture (based on our numerically measured low velocity fraction
 of planetary encounters) is
 \begin{eqnarray}
 \label{eqn:pcap}
 P_{cap} (s_1, R_{td}) &\sim & 0.2 \left. {N(R_{td}) \over N_c} \right|_{V_\infty=0}^{0.05} \left({R_{td} \over 0.1 }\right)^{-{1 \over 2}} \left({ M_p/M_* \over 10^{-3} }\right)^{1\over 6} \nonumber \\
&& 	\times 
	\left({\rho \over 1 ~{\rm g~cm}^{-3}} \right)^{1\over 6} 
	\left({ M_* \over M_\odot }\right)^{-{1 \over 3}} 
	\nonumber \\
&& \times \left({ s_1 \over 100 ~{\rm km}} \right)^{1 \over 2} 
   (1 + q_b)^{-{1 \over 3}}.
 \end{eqnarray}
 To estimate the capture probability we have used a fixed range for $V_\infty$ and corrected for
 the width of this range (rather than
  integrating to the limit given in equation \ref{eqn:vinfvk2}) so that at least a few dozen particles
 are counted at low $V_\infty$ for all planets in all integrations.   As the velocity distribution decreases
 with increasing $V_\infty$ for the outer planet and vice versa for the inner planet (see Figure \ref{fig:vhist}) 
 we have somewhat (but not significantly) underestimated the capture probability for the inner planet 
 and overestimated it for the inner planet.
 
\begin{figure}
\includegraphics[width=9cm]{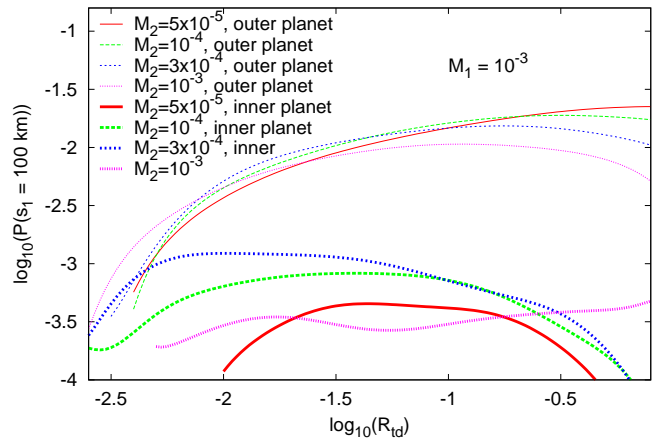}
\caption{The probability that a binary with primary radius $s_1 = 100 $~km would leave its secondary captured 
as a function of normalized tidal disruption radius, $R_{td}$,  for the integrations shown 
in Figures \ref{fig:qhist} and \ref{fig:v0hist}.  
Using equation (\ref{eqn:pcap}) the capture probably can be adjusted 
for different binary planetesimal masses, mass ratios and densities.   
The probabilities have been calculated from the fraction of integrated planetesimals approaching
planets at low velocities and using factors given in equation (\ref{eqn:fac}).
Thick lines correspond to probabilities for the inner planets and thin lines to those for the outer planets.
In all cases  the outer planet is more likely to capture satellites than the inner planet. 
The probability that the outer planet captures a 100 km radius planetesimal from a binary
with a tidal disruption radius $r_{td}/r_{Hp}$ in the range 0.01 to 0.1 is about 1/100.  
}
\label{fig:v0hist_s}
\end{figure}

 We show in Figure \ref{fig:v0hist_s} the probability of 
 capture of the secondary by each planet, for a primary planetesimal radius 
 $s_1=100$~km and low binary mass ratio, $q_b$, and for the same simulations shown in 
 Figure \ref{fig:qhist} and \ref{fig:v0hist}.  
 For other values of stellar mass, $M_*$, planetesimal density, $\rho$, 
 binary mass ratio, $q_b$, and planetesimal primary
 radius, $s_1$, the probability 
 can be adjusted using scaling given in equation (\ref{eqn:pcap}).
 
A comparison between Figures \ref{fig:v0hist} and \ref{fig:v0hist_s} shows that the probability of
capture is reduced from that computed with a fixed velocity range
for more weakly bound binaries (larger $R_{td}$).  This follows as
more weakly bound binaries have lower internal rotational velocities and so cannot exchange as
much kinetic energy during disruption as more tightly bound binaries.  We see from Figure \ref{fig:v0hist_s} 
that the probability of capture about the outer planet  is only weakly dependent on the mass
of the outer planet.  However,   
the contrast between the capture probability for the outer and inner planet  is higher when the planet mass ratio
is larger.    When the ratio of the planet masses $M_2/M_1$ is low, the capture probability for weakly bound
binaries (large $R_{td}$) is higher than for $M_2/M_1 \sim 1$.  This is counter-intuitive as it suggests that
additional weakly bound binaries are captured by the inner planet when the outer planet is more massive. 
The probability of capture by the outer planet is remarkably insensitive to the ratio of the planet masses. 
However it would be non-trivial to predict the shape of these probability curves and how they depend
on planet masses and ratios.
 
Our estimated capture probability $\sim 1\%$ is consistent with that estimated by \citet{nogueira11} for
the capture of Triton via binary exchange reactions during Neptune's slow migration outwards
following close encounters by the giant planets within the context of the ``Nice'' model.  But as shown here
only the outermost planet would tend to capture satellites during this slow period of migration, 
so other scenarios are required
to account for the similarity of the irregular satellite populations of the giant planets \citep{jewitt07}
and so the majority of irregular satellites in our Solar system.

\begin{figure}
\includegraphics[width=9cm]{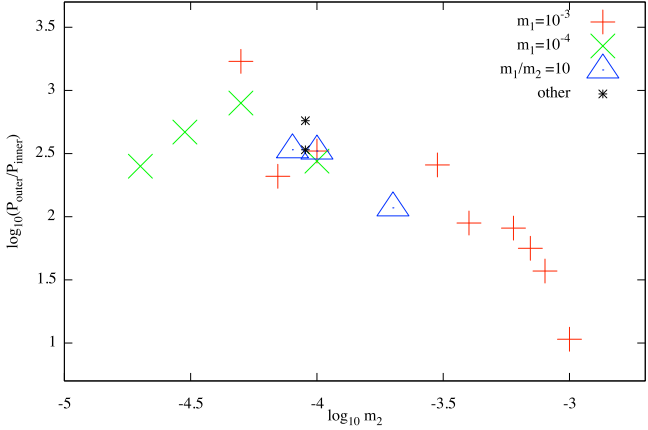}
\caption{The ratio of the satellite capture probability of the outer planet divided by that of the inner planet
($y$-axis) is shown as a function
of outer planet mass ($x$-axis) on a log scale.  Points for each simulation are taken from Table \ref{tab:tab1}.
For most situations the ratio of capture probabilities is two orders of magnitude suggesting that
outer planets would preferentially be detected from dust associated with an irregular satellite population.
}
\label{fig:sum_nov}
\end{figure}

If dust associated with a satellite population is detected in the vicinity of an outer exoplanet
but there is no inner planet detected, then it would be helpful if the ratio of satellite capture probabilies
could be used to place limits on the planet masses. 
In Figure \ref{fig:sum_nov} using different simulations 
we plot the ratio of the capture probability for the outer planet and that of 
the inner planet as a function  
of the outer planet's mass.  The capture probabilities are those at a disruption radius
 $R_{td}=0.1$ and for primary planetesimal radius  $s_1 = 100~$km and
 are  listed for the different simulations in Table \ref{tab:tab1}.
 Figure \ref{fig:sum_nov} shows that the difference in capture probabilities is sensitive to the outer
 planet mass with lower mass outer planets preferentially capturing more bodies.  However,
 the ratio of capture probabilities appears to be fairly insensitive to
 the inner planet mass.     We note that the two planet masses  were
 necessarily constrained to be in a narrow range because of numerical limitations 
 (we did not want to integrate more than a few thousand orbits).   
 More diverse and challenging simulations would be needed to
 determine if constraints on planet masses could be based on captured satellite populations.
 For the range covered by our simulations the ratio is about two orders of magnitude suggesting
 that outer planets would preferentially be detected from dust associated with an irregular satellite population.
 
\section{Collisional evolution of a population of captured irregular satellites}

Planetesimals captured into planetary orbit contribute to a dust producing collisional cascade
if they collide with other irregular satellites or other objects passing within the Hill radius of the planet.
\citet{kennedy11}  primarily considered the evolution of  a primordial irregular satellite population. 
However if a planet migrates outwards then the irregular satellite population can be replenished
by recently captured satellites.    The proximity of Fomalhaut's dust disk to Fomalhaut b suggests
that this planet could be migrating outwards. If so the planetesimal disk can provide a continuing source
of irregular captured satellites around Fomalhaut b.
The rate that irregular satellites are captured can be 
described as a mass rate, $\dot M_s$, the mass in irregular satellites captured per unit time.
The total rate is integrated over the planetesimal distribution,
however because the planetesimal mass distribution is likely dominated by the most massive objects
it is simpler to approximate the rate by only considering the massive binary population; 
\begin{equation}
\dot M_s \sim \dot m_o f_B q_{typ} P_{typ} 
\label{eqn:dotms}
\end{equation}
where $f_B$ is the fraction
of mass in massive binaries with typical separations, $a_{B,typ}$, mass ratios, $q_{typ}$,
and typical total mass, $m_{typ}$.  Here $P_{typ}$ is the probability that a binary with these
properties would be disrupted and leave a bound satellite around the outer planet.
The mass rate $\dot m_o$ is the mass of planetesimals from the disk that 
is launched per unit time into orbits that cross the planets.  This rate is determined by the 
migration rate of the outer planet or 
\begin{equation}
\dot m_o = 2 \pi a_p \Sigma_d \dot a_p \sim M_{disk} {a_p \over w_{disk} } {\dot a_p \over a_p}
\label{eqn:dotmo}
\end{equation}
where $a_p$ is the semi-major axis
of the outer planet that is migrating at a rate $\dot a_p$ into a planetesimal disk of surface density $\Sigma_d$.
The planetesimal disk has mass $M_{disk} \sim 2 \pi a_p w_{disk} \Sigma_d$ 
where $w_{disk}$ is the width of the planetesimal belt.  
In the case of a single planet embedded in a disk  \citep{ida00,kirsh09},
when there is adequate disk material to maintain migration
then the migration rate is independent of the planet mass with 
\begin{equation}
{\dot a_{p,single} P \over a_p} \approx {4 \pi \Sigma a_p^2 \over M_*} 
\sim 2{M_{disk} \over M_*} {a_p \over w_{disk} }
\label{eqn:dotap}
\end{equation}
where $P$ is the orbital period of the planet.
We can describe planet migration rates in terms of this fiducial value and using a unitless 
parameter $\eta_{mig}$ with
 $\dot a_p \equiv \eta_{mig} \dot a_{p,single}$.
For comparison, in our integrations the outer planets have migration rates ranging from $\eta_{mig} \sim$ 0.2 -- 1.5.

Combining equations (\ref{eqn:dotms}, \ref{eqn:dotmo}, \ref{eqn:dotap}) we find
\begin{eqnarray}
\label{eqn:dotms2}
{\dot M_s P \over M_p} &\sim& \left(M_{disk} \over M_p \right)^2 \left(a_p \over w_{disk}  \right)^{2} 
\left(M_p \over M_* \right)2 \eta_{mig} q_{typ} f_B P_{typ} \nonumber  \\
&\sim&  10^{-6} \left({ M_{disk} / M_p \over 0.1} \right)^2 \left( {w_{disk}/ a_p \over 0.2} \right)^{-2} 
\\ 
&& \times \left({ M_p /M_* \over 10^{-3} }\right) 
 \eta_{mig} q_{typ} \left({f_B \over 0.3}\right) \left({ P_{typ} \over 0.01} \right). \nonumber
\end{eqnarray}
where we have chosen the fraction of mass in binaries, $f_B$, appropriate 
for the low inclination transneptunian objects   \citep{noll08}.
Because the capture probability depends on the binary velocity (that has only a narrow range)
low mass objects are unlikely to be captured.  The above mass rate then represents few and rare
events occurring over a moderate time period.
As discussed by previous studies
\citep{estrada06,philpott10,bottke10} a system with only a few irregular satellites could still experience
significant collisional evolution.

We can consider a steady state where the captured massive irregular satellites drive a collisional cascade 
in the vicinity of the planet with a dust production and mass loss rate
$\dot M_d$ equivalent to the rate that mass is gained into the planet's Hill radius via capture of satellites; 
$\dot M_d = \dot M_s$.
 
The dust production rate can be estimated from the opacity of the small dust particles in the system, $\tau_d$
with radius $s_d \sim 10 \mu$m. Smaller dust particles are removed from the system due to radiation pressure
 or PR drag (see discussion by \citealt{kennedy11}).
The rate of collisions is $\sim \tau_d \Omega$ and so the dust production rate
\begin{equation}
\dot M_d \sim \rho_d s_d r_c^2 \tau_d^2 \Omega(r_c) 
\end{equation}
where $r_c$ the radius of the dust cloud and the dust has density $\rho_d$.
The luminosity of the dust cloud in units of the stellar luminosity depends on the opacity and size
of the cloud
\begin{equation}
{L_d \over L_*} \sim {\tau_d r_c^2 Q_a \over 4 a_p^2}
\end{equation}
where $Q_a$ is the albedo.
The luminosity ratio is an observable quantity and so relevant for the discussion
of emission from Fomalhaut b.

Inside the Hill radius of a planet we can describe radii in units of the Hill radius $\xi \equiv r/r_H$.
The angular rotation rate $\Omega(r)/\Omega_K = \xi^{-3/2}$ where $\Omega_K$ is that of the planet
in orbit about the star.  
We define $\xi_c \equiv r_c/r_H$ as the irregular satellite cloud radius in units of the planet's Hill radius.
The luminosity ratio and dust production rate can be written
\begin{eqnarray}
\dot M_d &\sim &\rho_d s_d \tau_d^2 \xi_c^{1/2} \mu^{2/3} a_p^2 \Omega_K \\
&\sim & 3 \times 10^{-4} {M_\oplus \over {\rm Myr}}
\left(  { \xi_c \over 0.2 } \right)^{{1 \over 2}}  
\left(  { M_p/M_* \over 10^{-3}} \right)^{{2\over 3}}
\left({\tau_d \over 0.01}\right)^{2}
 \nonumber \\
&& \times 
\left(  {  \rho_d \over 1 {\rm g~cm}^{-3} } \right)
\left(  {  s_d \over 10 \mu{\rm m}} \right)
\left(  { a_p \over 10 AU} \right)^{1 \over 2}
\left(  {  M_*\over M_\odot} \right)^{1\over 2}
\nonumber
\end{eqnarray}
where $\mu = M_p/M_*$ and
\begin{eqnarray}
{L_d \over L_*} & \sim & {\tau_d \xi_c^2 \mu^{2/3} Q_a \over 4 \times 3^{2/3}} \\
& \sim & 5 \times 10^{-8}
\left({\tau_d \over 0.01}\right)
\left(  { \xi_c \over 0.2 } \right)^{{ 2}}  
\left(  { M_p/M_* \over 10^{-3}} \right)^{{2\over 3}}
\left( { Q_a \over 0.1}\right).
 \nonumber
\end{eqnarray}
Here we have chosen parameters ($\tau_d, Q_a, \xi_c, M_p$) based on previous scenarios that account
for Fomalhaut b's emission \citep{kalas08,kennedy11}.

So as to compare to a planet migration rate
we show the dust production rate in units of the planet's mass and orbital period, $P$,
\begin{eqnarray}
\label{eqn:dotmd2}
{\dot M_d  \over M_pP^{-1}} &\sim & 2 \pi \rho_d s_d \tau_d^2 \xi_c^{1/2} \mu^{2/3} a_p^2 M_p^{-1} \\
&\sim & 3 \times 10^{-9}
\left(  { \xi_c \over 0.2 } \right)^{{1 \over 2}}  
\left(  { M_p/M_* \over 10^{-3}} \right)^{-{1\over 3}}
\left({\tau_d \over 0.01}\right)^{2}
 \nonumber \\
&& 
\left(  {  \rho_d \over 1 {\rm g~cm}^{-3} } \right)
\left(  {  s_d \over 10 \mu{\rm m}} \right)
\left(  { a_p \over 100 AU} \right)^{ 2}
\left(  {  M_*\over M_\odot} \right)^{-{2 \over 3}}.
\nonumber
\end{eqnarray}
We can compare the rate that mass is captured into planetary orbit (equation \ref{eqn:dotms2}) 
with the dust production 
rate (equation \ref{eqn:dotmd2}).   
The above extremely modest dust production rate is below the mass that
could be captured in irregular satellites.   
A collisional cascade could be maintained if even a small fraction of orbit crossing planetesimals 
are captured into an irregular satellite population as the outer planet migrates outwards.
If the collision rate and opacity becomes high then small objects
could be re-accreted onto larger ones.  
At  a sufficiently high opacity a dense and thin ring system could form.

As discussed by \citet{kennedy11} there is only a small range of parameter space for
possible irregular satellite clouds
that could account for the optical emission in the vicinity of Fomalhaut b.   Here we find that
this constraint might
relaxed if the cloud is replenished by incoming captured satellites.    
 These would have to be fairly massive (and so rare) to be captured, however their
lifetime depends not only the timescale that they collide with other object bound objects but also with
the time it takes them to 
collide with the population of orbit crossing planetesimals (including planetesimals that are not binaries).  
As the planet migrates outwards the irregular satellite population would build up until there is a balance
between mass acquired by the system (in the form of captured objects that are then fragmented
by collisions with either each other or external objects) and mass lost 
that is either removed from the system by radiation forces, or accreted back onto other objects.
If the opacity is sufficiently high then debris could coalesce into a ring.

\section{Discussion and Summary}

Using numerical integrations of planets interacting with a planetesimal belt we have explored
the statistics of close encounters of planetesimals with planets during planetary migration.
We have focused exclusively on the problem of a two planet system in proximity to a cold (in terms
of velocity dispersion) planetesimal belt  exterior to both planets that induces planetary migration.
If the planetesimal belt contains binary planetesimals then close encounters with planets can
disrupt these binaries.   To predict the location of tidal disruption (at what radius and which planet),
 we have taken into account the order of close encounters with planets.
We find that
the velocity distribution (for velocities with respect to the planet) of tidally disrupting encounters 
differs for the two planets.  Lower velocity encounters are only likely with the outer planet.    
As a consequence we predict that the probability that irregular satellites are captured due to binary exchange reactions is higher, by about two orders of magnitude, 
with an outer planet than an inner one.   We estimate
that planetesimal binaries similar to those in the Kuiper belt would have a probability of about 1/100 of disrupting
and leaving a bound object about an outer migrating planet.  Our numerical estimate is consistent with that
recently estimated by \citet{nogueira11} for the capture of Triton into orbit about Neptune.
In migrating systems we infer that only the outer planet could gain a significant population
of captured irregular satellites and other scenarios would be needed to account for irregular satellite populations
around inner giant planets.  We note that the simulations considered here were 
restricted by numerical considerations
to a narrow range of planet and disk masses, only considered cold and exterior planetesimal disks, only considered
two giant planet systems and situations lacking planetary encounters.   Future studies can expand the types of
systems numerically integrated to predict the probability of satellite capture for more diverse systems.

We estimate that the tidal disruption radius in units of the planet's Hill
radius is approximately set by the binary separation in units of its own Hill radius.  Thus if optical emission
from exoplanets is associated with a dust cloud generated by irregular satellites, 
as explored by \citet{kennedy11}, then
there is a relation between the size of the dust cloud, the planet mass and the properties of the primordial
binary planetesimal population. 
We estimate that capture is allowed only if the incoming binary velocity, with 
respect to the planet is only a few times the velocity of the binary orbit prior to disruption. 
We have used this estimate and
 our numerically measured encounter distributions to estimate the probability that the secondary of
a binary planetesimal could be captured into orbit about a planet.   To improve
upon our derived capture probabilities it is likely that 
the four body problem (binary planetesimal encounter with a planet in orbit about a star) must be integrated. 

\citet{kennedy11} suggested that irregular satellites can be captured if Fomalhaut b crosses into
the planetesimal belt.  Here we explore a related scenario. We consider  Fomalhaut b not crossing into
the planetesimal belt, but  migrating
outwards and capturing planetesimals due to binary exchange reactions with binaries that
have been perturbed into planet orbit crossing trajectories.  A crude estimate of the 
mass in irregular satellites captured per unit time suggests that the mass captured per unit time 
could be larger than  
the estimated dust production rate if a collisional cascade operates among the irregular satellite population.
Consequently there could be sufficient mass captured as irregular satellites to fuel such a 
collisional cascade.    In a migration scenario, we find that the outer planet is likely to gain two orders
of magnitude more irregular satellites than inner planets.  It is tempting to consider this scenario as
accounting for the detection of Fomalhaut b but not other, possibly more massive planets that 
could exist interior to  it in the system \citep{kenworthy09}.
Further study is needed
to determine whether a migration scenario for Fomalhaut b that included irregular satellite capture would
be consistent with the $\sim 0.1$ eccentricity of the dust ring and planet and 
the location and shape of the ring edge.


\vskip 0.3 truein
Acknowledgements.   This work was in part supported by NSF through award AST-0907841.   
We thank NVIDIA for gift of graphics cards.  We thank Scott Gaudi for interesting comments.


\begin{thebibliography}{}

\bibitem[Agnor \& Hamilton(2006)]{agnor06}
Agnor, C. B. \& Hamilton, D. P. 2006, Nature, 441, 192

\bibitem[Bottke et al.(2010)]{bottke10}
Bottke, W. F., Nesvorny, D., Vokrouhlicky, D., \&  Morbidelli, A.  2010, AJ, 139, 994

 \bibitem[Canup \& Ward(2002)]{canup02} 
Canup, R. M.. \& Ward, W. R.	2002, AJ, 124, 3404	

\bibitem[Estrada \& Mosqueira(2006)]{estrada06}
Estrada, P. R., \& Mosqueira, I. 2006, Icarus, 181, 496


\bibitem[Fabrycky \& Murray-Clay(2010)]{fabrycky10}
Fabrycky, D. C., \& Murray-Clay, R. A. 2010, ApJ, 710, 1408 

\bibitem[Fernandez \& Ip(1984)]{fip84} 	     
Fernandez, J. A., \& Ip, W.-H.	1984 , Icarus, 58, 109	     


\bibitem[Gozdziewski \& Migaszewski(2009)]{god09}
Gozdziewski, K., \& Migaszewski, C.  2009, MNRAS, 397, 16 

 \bibitem[Grundy et al.(2011)]{grundy11}
Grundy, W. M., Noll, K. S., Nimmo, F., Roe, H. G., Buie, M. W., Porter, S. B., Benecchi, S. D., Stephens, D. C., Levison, H. F., \& Stansberry, J. A.	2011, Icarus, 213, 678	


\bibitem[Gould \& Quillen(2003)]{gould03}
Gould, A., \& Quillen, A. C. 2003, ApJ, 592, 935


\bibitem[Ida et al.(2000)]{ida00}
Ida, S., Bryden, G., Lin, D.N.C., \& Tanaka, H., 2000, ApJ,   534, 428

\bibitem[Jewitt \& Haghighipour(2007)]{jewitt07}
Jewitt, D., \&   Haghighipour, N. 2007, Annu. Rev. Astron. Astrophys., 45, 261

\bibitem[Kalas et al.(2005)]{kalas05}
Kalas, P., Graham, J. R. \& Clampin, M. 2005, Nature 435, 1067

\bibitem[Kalas et al.(2008)]{kalas08}
Kalas, P. et al. 2008, Science, 322, 1345 

\bibitem[Kennedy \& Wyatt(2011)]{kennedy11}
Kennedy, G. M., \& Wyatt, M. C. 2011, MNRAS, 412, 2137	

 \bibitem[Kenworthy et al.(2009)]{kenworthy09}
Kenworthy, M. A., Mamajek, E. E., Hinz, P. M., Meyer, M. R., Heinze, A. N., Miller, D. L.,  Sivanandam, S., 
\& Freed, M.	2009, ApJ, 697, 1928

\bibitem[Kirsh et al.(2009)]{kirsh09}
Kirsh, D.R., Duncan, M., Brasser, R., \& Levison, H.F. 2009, Icarus, 199, 197

\bibitem[Levison \& Duncan(1997)]{levison97}
Levison, H. L., \& Duncan, M. J. 1997, Icarus, 127, 13

\bibitem[Marsh et al.(2005)]{marsh05}	
Marsh, K. A., Velusamy, T., Dowell, C. D., Grogan, K., \& Beichman, C. A.	2005ApJ...620L..47M

\bibitem[Marois et al.(2010)]{marois10}
Marois, C., Zuckerman, B., Konopacky, Q. M., MacIntosh, B., \&  Barman, T. 2010,
Nature, 468, 7327, 1080

\bibitem[Moore \& Quillen(2011)]{moore11}
Moore, A., \& Quillen, A. C. 2011, New Astronomy, 16, 445

\bibitem[Nesvorny et al.(2007)]{nes07}
Nesvorny, D., Vokrouhlicky, D.,  \& Morbidelli, A.  2007, ApJ, 133, 1962

\bibitem[Noll et al.(2008)]{noll08}
Noll, K.S., Grundy, W.M., Stephens, D.C., Levison, H.F., Kern, S.D., 2008, Icarus, 194, 758

\bibitem[Nogueira et al.(2011)]{nogueira11}
Nogueira, E., Brasser, R., \& Gomes, R. 2011, Icarus, 214, 113

\bibitem[Philpott et al.(2010)]{philpott10}
Philpott, C. M., Hamilton, D. P., \& Agnor, C. B. 2010, Icarus, 208, 824


\bibitem[Quillen et al.(2007)]{quillen07}
Quillen, A. C., Morbidelli, A., \& Moore, A.	2007, MNRAS, 380, 1642




\bibitem[Su et al.(2009)]{su09}
 Su, K. Y. L., Rieke, G. H., Stapelfeldt, K. R., Malhotra, R., Bryden, G., Smith, P. S., Misselt, K. A., Moro-Martin, A., 
\& Williams, J. P. 2009, ApJ, 705, 314	

\bibitem[Tsiganis et al.(2005)]{nice}
Tsiganis, K., Gomes, R., Morbidelli, A., \& Levison, H. F. 2005, Nature, 435, 459

\bibitem[Vokrouhlicky et al.(2008)]{vok08}
Vokrouhlicky, D., Nesvorny, D.  \& and  Levison, H. F. 2008, AJ, 136, 1463

\end{thebibliography}
\end{document}